\documentclass[twocolumn]{aastex631}
\usepackage{comment}
\usepackage{natbib} 
\usepackage{amsmath}
\usepackage[flushleft]{threeparttable}
\usepackage{hyperref}

\newcommand{\kms}{km s$^{-1}$}

\newcommand{\hii}{H~{\scriptsize II}}

\begin{document}

\author[0009-0005-4373-0200]{Nora Salem}
\affiliation{Depts of Physics \& Astronomy, Haverford College, 370 Lancaster Ave, Haverford, PA 19041, USA}

\author[0000-0002-5811-0136]{Dylan M. Par\'e}
\affiliation{Joint ALMA Observatory, Alonso de Cordova 3107, Vitacura, Casilla 19001, Santiago de Chile, Chile}
\affiliation{National Radio Astronomy Observatory, 520 Edgemont Road, Charlottesville, VA 22903, USA}

\author{Paulo Cortes}
\affiliation{Joint ALMA Observatory, Alonso de Cordova 3107, Vitacura, Casilla 19001, Santiago de Chile, Chile}

\author[0000-0002-6753-2066]{Mark R. Morris}
\affiliation{Department of Physics \& Astronomy, University of California, Los Angeles, 475 Portola Pl., Los Angeles, CA 90095-1547, USA}

\author[0000-0002-5714-799X]{Valentin J. M. Le Gouellec}
\affiliation{Institut de Cienci\`es de l’Espai (ICE-CSIC), Campus UAB, Carrer de Can Magrans S/N, E-08193 Cerdanyola del Vall\`es, Spain}
\affiliation{Institut d’Estudis Espacials de Catalunya (IEEC), c/ Gran Capitá, 2-4, 08034 Barcelona, Spain}

\title[ALMA RA]{\uppercase{The Polarization and Magnetic Field of the Radio Arc as Observed by ALMA at 100 GHz}}

\correspondingauthor{Nora Salem}
\email{nsalem@haverford.edu}

\begin{abstract}

The unique Galactic Center non-thermal filaments (NTFs) have been a focus of investigations for over 40 years. The most prominent manifestation of the NTFs is a bundle of parallel filaments known as the Radio Arc. Radio polarimetric observations made with the Very Large Array (VLA) at 10 GHz have revealed an alternating magnetic field pattern in the Radio Arc that could either be a result of multiple field systems being encountered along the line of sight or an intrinsic feature of the Radio Arc. These VLA observations were not able to distinguish between these possibilities due to the large rotation measures encountered towards the source. We present ALMA 100 GHz observations of the Radio Arc that are not impacted by significant Faraday effects. The observations reported here represent both the first time that ALMA has been used to study the NTFs and the first time 100 GHz polarimetric observations have been conducted on the Radio Arc. We find a uniformly rotated magnetic field with respect to the NTF filament orientation, with the angle of rotation being constant along the length of each filament. However, we find a systematically different magnetic field orientation in different Radio Arc filaments. We use this field pattern to update our understanding of the line-of-sight structures local to the Radio Arc. We find that the magnetic field inferred from our ALMA observations is likely a result either of confusion from multiple magnetic field systems or because the polarization is centrally concentrated within the NTF filaments.
 
\end{abstract}



\section{INTRODUCTION} \label{sec:intro}
The Galactic Center (GC) hosts an extreme environment, with higher densities \citep[10$^3$-10$^6$ cm$^{-3}$,] []{Mills2018}, temperatures \citep[50-300 K,][]{Krieger2017}, magnetic field strengths \citep[100s $\mu$G-10s mG, e.g.,][]{Yusef-Zadeh1987,Chuss2003a, Pillai2015, Guerra2023}, and more turbulent velocities \citep[$\sim10$ \kms\,][]{Kauffmann2017} than in the Galactic disk. The GC is the closest galactic nuclear region to Earth at $\sim$8.2 kpc \citep{Abuter2019}, making it the best available laboratory to study such an environment. Assuming that the GC is representative of other galactic centers, we can infer the properties of extragalactic nuclear regions that are too distant to be resolved with current instruments.

Radio observations towards the GC have revealed a population of thin, glowing strands with lengths up to 10s of parsecs and thicknesses of only a fraction of a parsec \citep[e.g.,][]{YMC1984, Yusef-Zadeh2022, Yusef-Zadeh1986a, Morris1996a, Gray1995, Pare2019, Pare2021, Pare2022, Pare2024b}. Due to the non-thermal nature of these structures they have become known as the non-thermal filaments (NTFs). These NTFs are highly polarized sources (percentage polarizations $\geq$50\%) illuminated by synchrotron-emitting relativistic electrons. The magnetic field internal to these structures is generally oriented parallel to the NTFs \citep{YWP1997,Lang1999a,Lang1999b}. Since the NTFs are generally oriented perpendicular to the Galactic plane, this suggests a vertical magnetic field that pervades the GC \citep{YMC1984, Yusef-Zadeh1987, LaRosa2006,Morris2006sum,Morris2015}. The Radio Arc is one of the brightest systems of NTFs, and was also the first discovered NTF and therefore has been studied across multiple wavelengths at increasingly higher resolution \citep[e.g.,][]{YMC1984,Inoue1989,Pare2019}. 

More recent observations of NTFs have begun to reveal examples where the NTF magnetic field orientation is apparently rotated or even perpendicular to the orientation of the filaments. Recent Very Large Array (VLA) observations at 10 GHz of the Radio Arc and other NTFs have shown alternating magnetic fields that regularly transition from parallel to rotated with respect to the filament orientation \citep{Pare2019,Pare2024b}. These rotated field regions may be an intrinsic feature of the Radio Arc or may be caused by interactions with local and foreground magnetized structures \citep{Simpson2007,Butterfield2018}. It is also possible that intervening Faraday rotation effects have not been fully corrected \citep{Pare2021}. The VLA observations were not able to distinguish between these possible scenarios because of the significance of Faraday effects at 10 GHz. 

The 100 GHz frequency of the Atacama Large Millimeter/submillimeter Array (ALMA) observations reported here are much less impacted by Faraday effects because of the $\lambda^2$ dependence of the Faraday effect, allowing us to refine our understanding of whether the orientation offset of the inferred magnetic field with respect to the NTF orientation is a spurious Faraday rotation effect. The Radio Arc is known to be bright at 100 GHz frequencies, as shown from previous 150 GHz single-dish observations made using the Nobeyama 45-m telescope \citep{Reich2000}. The Radio Arc is also visible in 2 mm 30-m IRAM observations \citep{Staguhn2019}, although neither of these observations have polarimetry. Previous polarimetric observations at $\sim$90 GHz also exist for the Radio Arc as obtained using ACTpol \citep{Guan2021}, but with a resolution of 1\arcmin. The polarimetric aspect of the ALMA observations presented here therefore represent the first time that the magnetic field for this source has been studied at such a high frequency at sub-arcsecond resolution.

In Section \ref{sec:meth} we present the details of the ALMA observations, the calibration methods, and the imaging parameters used to study the Radio Arc. In Section \ref{sec:res} we present the total and polarized intensity distributions obtained for the Radio Arc as well as the derived magnetic field distribution. In Section \ref{sec:disc} we describe how these ALMA results expand on previous polarimetric studies of the Radio Arc. We also develop a sketch of the geometry local to the Radio Arc to explain the proposed magnetic field distribution obtained from our ALMA observations. We conclude the paper in Section \ref{sec:conc}.

\section{Observations and Data Reduction} \label{sec:meth}

\subsection{Observations} \label{sec:obs}
\begin{deluxetable*}{lcccccccc}
\tablecaption{ALMA Observation Parameters
\label{tab:alma_obs}}
\tablewidth{0pt}
\tablehead{
\colhead{Band \# ($\nu$)} & \colhead{Cent. R.A.} & \colhead{Cent. Decl.} &  \colhead{Date} & \colhead{Res.} & \colhead{Array Conf.} & \colhead{Int. Time (min)} & \colhead{\# pointings} & \colhead{Mosaic Size}}
\startdata
Band 3 (104.5 GHz) & 17:46:34.9 & -28.51.10.3 & 03-03-2025 & 1.5\arcsec & C43-3 & 4 & 150 & 900\arcsec\ $\times$ 270\arcsec
\enddata
\tablecomments{The first column indicates the ALMA band and central frequency used for the observations. The second and third columns indicate the central right ascension and declination used as the phase center coordinates for the creation of the images shown in this letter. The fourth column indicates the date when the source was observed by ALMA. The fifth column indicates the angular resolution of the ALMA observations. The sixth column displays the ALMA array configuration used to obtain these observations. The seventh column indicates the total on-source observation time obtained per pointing for the observations of the target. The eighth column indicates how many ALMA pointings were used to create the mosaic images presented in this letter, and the ninth column indicates the size of the resulting mosaic in arcseconds.}
\end{deluxetable*}
We observed the Radio Arc in 150 pointings using ALMA Band 3 with a central wavelength of 104.5 GHz (project code: 2024.1.00515.S). The resulting mosaic has a resolution of 1.5\arcsec\ with a pixel size of 0.3\arcsec\ (for 5 pixels across each beam) and the total observation time on-source was $\sim$3 hours (three $\sim$1-hour sessions), resulting in a sensitivity level of $\sim$100 $\mu$Jy. These ALMA observations have baseline lengths ranging from 15 to 260 meters. More details regarding the observations are displayed in Table \ref{tab:alma_obs}.

We note that additional observation time was obtained ($\sim$7 hours on-source) but the polarization calibrator was not observed for long enough in these other execution blocks to obtain sufficient parallactic angle coverage. The lack of sufficient parallactic angle coverage resulted in these execution blocks being marked as semi-pass in the ALMA quality check process. We therefore chose to not include these execution blocks for the results presented here. 

\subsection{Calibration} \label{sec:cal}
J1733-1304 was used as the flux and bandpass calibrator, J1924-2914 was used as the polarization calibrator, and J1744-3116 was used as the phase calibrator. The bandpass and polarization calibrators were observed once at the beginning of each observation block for 5 minutes each in each execution block, and J1744-3116 was observed for 30 seconds every $\sim$8 minutes between the target observations. 

The data reduction was done using the Common Astronomy Software Application (CASA) utilizing the ALMA calibration pipeline (CASA pipeline version 6.6.6.17), which now includes polarization calibration. The pipeline calibrated products were of sufficiently high quality to not require any additional manual calibration. We verified that J1924-2914 was observed to be significantly polarized by imaging the polarization calibrator after applying the pipeline calibration to the data set. We obtained a Stokes $I$ intensity of $\sim$5 Jy beam$^{-1}$ and Stokes $Q$ and $U$ intensities of $\sim$0.1 Jy beam$^{-1}$, equating to a percentage polarization of $\sim$2\% for J1924-2914. This Stokes $I$ value and percentage polarization agree with previous ALMA observations of this calibrator \citep[e.g.,][]{Goddi2021}.

\subsection{Imaging} \label{sec:imag}
\begin{figure*}
    \centering
    \includegraphics[width=1.0\linewidth]{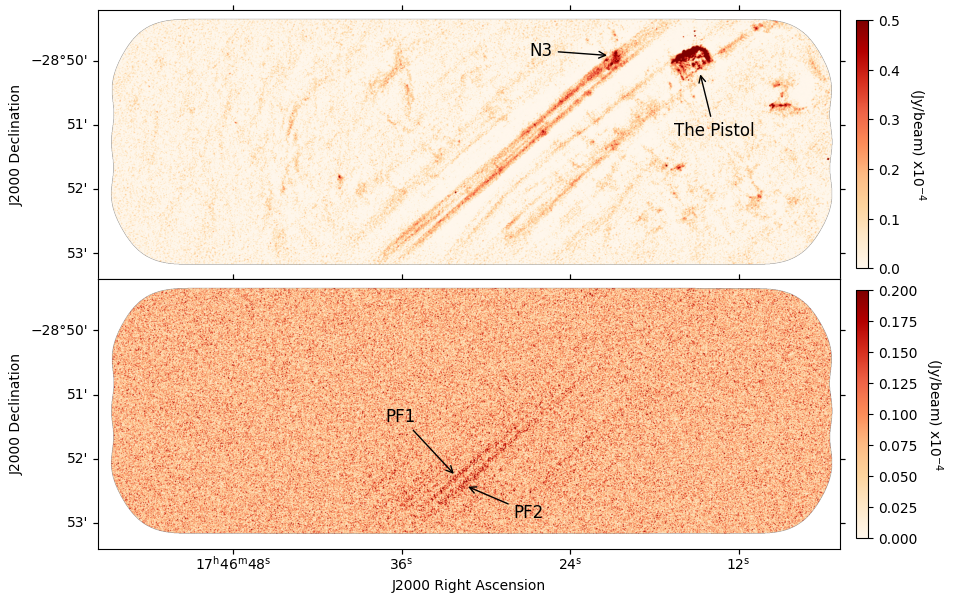}
    \caption{(top) Total intensity and (bottom) polarized intensity of the Radio Arc at 100 GHz. Arrows in the top panel point towards notable features that are not components of the Radio Arc in the field of view. The arrows in the bottom panel mark the two distinct polarized filaments in the Radio Arc that are discussed in detail in the text.}
    \label{fig:2panel}
\end{figure*}

The $\sim$100 GHz observations presented here represent the first time ALMA has been used to observe the unique NTF structures in the GC polarimetrically at such high resolution. Figure \ref{fig:2panel} displays the total intensity and polarized intensity of the Radio Arc obtained from ALMA using the calibration pipeline.  The top panel of Figure \ref{fig:2panel} marks the locations of the unusual N3 point source located within the NTFs \citep{Ludovici2016} and the nearby Pistol \hii\ region \citep{YMC1984,Figer1998,Figer1999pist}.

Though we used the CASA ALMA pipeline for the data calibration, we performed the imaging manually using the same CASA version as what was used for the pipeline calibration. The continuum total intensity image was made using the CASA \textit{tclean} task using mfs mode with 100,000 iterations. This imaging was performed using the mosaic gridder with an image size of 3000 $\times$ 900 pixels and a pixel size of 0.3\arcsec. We also used a primary beam threshold of 0.2 to de-emphasize the emission at the edge of the mosaic field of view where the ALMA observations are less sensitive. We used a briggs weighting with a robust parameter of 0.5 and a multiscale deconvolver with scales of 0, 5, and 15 $\times$ the ALMA beam size. We obtained separate Stokes \textit{Q} and \textit{U} using the same imaging parameters as described above but with only 1000 cleaning iterations. The \textit{Q} and \textit{U} images allow us to obtain the total polarized intensity and observed polarization angle via the following equations:

\begin{equation}
\label{eq1}
    P=\sqrt{Q^2+U^2},
\end{equation}

\begin{equation}
\label{eq2}
    \chi=\frac{1}{2}\text{tan}^{-1}\left(\frac{U}{Q}\right).
\end{equation}
The CASA task \textit{immath} performs Equations \ref{eq1} and \ref{eq2} on the cleaned \textit{Q} and \textit{U} images to produce \textit{P} and $\chi$ distributions. As seen in the lower panel of Figure \ref{fig:2panel}, there are two distinct filaments of higher polarized intensity which we label PF1 and PF2. The properties of these highly polarized filaments are studied in more detail later in this work.

Lastly, we obtain a debiased polarized intensity image by subtracting out the background polarized flux level from the total polarized intensity image. Debiasing helps accounts for weaker polarization in the edges of the beam and other instrumental defects. We choose a 2$\sigma$ cutoff of 1.61$\times 10^{-4}$ Jy/beam to subtract out the background noise while still highlighting the filaments. 

We also made spectro-polarimetric cubes with 4 channels, one channel for each of the four spectral windows comprising the ALMA observations. This spectral cube allowed for an assessment of the significance of potential Faraday effects by measuring the Rotation Measure (RM) towards the Radio Arc such that: $\chi = RM\times{}\lambda^2 + \chi_0$, where $\chi$ is the observed polarization angle (rad), $RM$ is the Rotation Measure observed towards the Radio Arc (rad m$^{-2}$), $\lambda$ is the wavelength of the observation, and $\chi_0$ is the intrinsic polarization angle emitted by the source. By fitting the spectrum of the polarization angle as a function of wavelength squared with a linear model for each line of sight we obtain the the RM as the slope of the best linear fit for each pixel. We found from this model fitting that RM effects are negligible ($\sim$10 rad m$^{-2} \pm$ 5 rad m$^{-2}$) at the 100 GHz ALMA frequencies of this data set for the lines of sight corresponding to the significant polarization studied. Rotation Measures of this magnitude will not significantly rotate the observed polarization angle at the wavelengths of this ALMA observation.

As a further check on the continuum mosaic obtained for the Radio Arc we also created separate images of each of the 150 ALMA pointings comprising the mosaic. For those pointings that overlap with significant polarization we verified that the same polarization features or obtained in the individual pointings as are seen in the full mosaic.

For the subsequent discussion we mask out the polarized intensity that lies spatially outside of the total intensity bounds of the Radio Arc. See Figure \ref{fig:debiasedp} for an image of the debiased and masked polarized intensity overlaid on the total intensity contours. 

\section{Results}  \label{sec:res}
\subsection{Debiased Polarized Intensity and the Derived Magnetic Field}

\begin{figure*}
    \centering
    \includegraphics[width=1.0\linewidth]{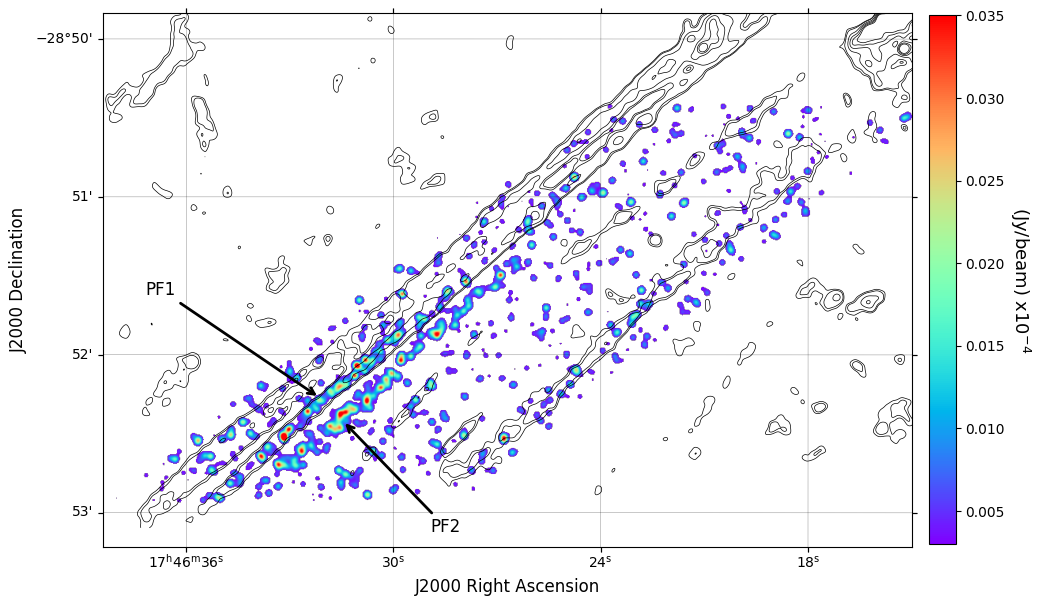}
    \caption{Debiased smoothed polarized intensity at 100 GHz where the polarization with values less than 2$\sigma$ (1.61 (Jy/beam)$\times 10^{-4}$) is masked. The image was smoothed with a Gaussian filter using a sigma of 3. Overlaid are contours of the total intensity distribution indicating the locations of the Radio Arc filaments and are shown at values of 0.7, 1.0, 2.0, 3.0, 4.0, 5.0 (Jy/beam)$\times 10^{-4}$. Features PF1 and PF2 are also indicated.}
    \label{fig:debiasedp}
\end{figure*}

In Figure \ref{fig:debiasedp} we present the continuum debiased and masked polarized intensity of the Radio Arc in colorscale with the total intensity contours shown as black lines. We observe the strongest intensities coinciding with the total intensity around RA=17:46:27 to RA=17:46:34. There are two distinct filaments of polarized intensity in this RA range which we refer to as ``Polarized Filaments'' (PFs), one of which (PF1) spatially corresponds to the total intensity contours, while the other one (PF2) is seemingly not associated with significant total intensity. 

We also observe significant polarized intensity only in the Southernmost extent of the Radio Arc filaments. The Northern portion of the Radio Arc exhibits no significant polarization. This depolarization effect has been observed in previous polarimetric observations of the Radio Arc, with the depolarization possibly being the result of a more complicated line-of-sight geometry in this region of the Radio Arc or the increased proximity with prominent \hii\ regions \citep{Yusef-Zadeh1987,Pare2019}.

\subsection{The Derived Radio Arc Magnetic Field}
\begin{figure*}
    \centering
    \includegraphics[width=1.0\linewidth]{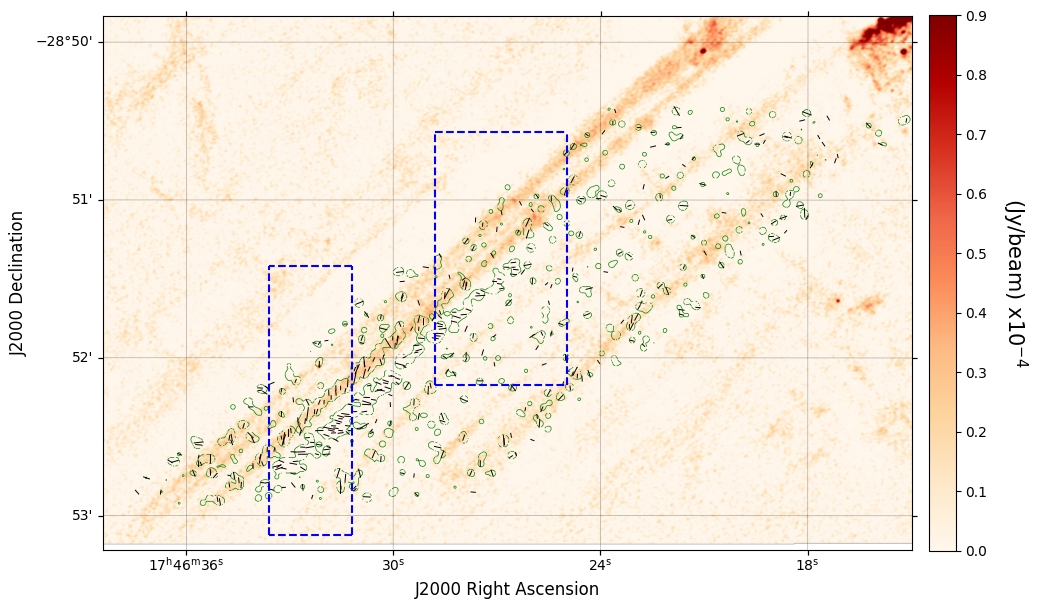}
    \includegraphics[width=1.0\linewidth]{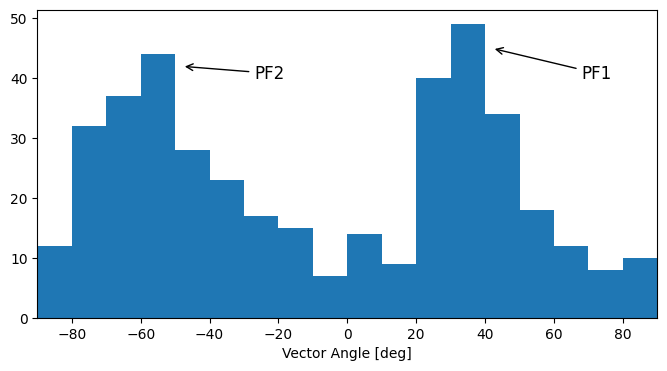}
    \caption{(top) Magnetic field vectors obtained by rotating the intrinsic polarization angles by 90$^{\circ}$, plotted over the 100 GHz total intensity. The green contour shows the smoothed debiased polarized intensity at 0.004 (Jy/beam)$\times 10^{-4}$. Blue dashed regions represent areas where the magnetic field is predominantly rotated with respect to the NTF orientation as observed by the VLA  \citep{Pare2019}. (bottom) Distribution of the ALMA magnetic field orientations shown in the upper panel wrapped from -90$^{\circ}$ to 90$^{\circ}$, with bin boundaries every 10$^{\circ}$. 0$^{\circ}$ corresponds to a magnetic field oriented parallel to the NTF orientation. Orientation enhancements corresponding to PF1 and PF2 are also marked. Not pictured: polarization angle uncertainties of $\pm0.29^{\circ}$}.
    \label{fig:bfield}
\end{figure*}

We determine the magnetic field orientation only where the debiased polarization is significant at the 2$\sigma$ level. Here debiasing refers to a correction for the positive bias of the polarization noise distribution, since polarization obeys a Rice distribution rather than a Gaussian distribution. We chose a 2$\sigma$ debias level since this corrects for the Ricean positive bias and removes most polarized emission that is not associated with the NTFs. For the lines of sight with significant polarization (polarization above the 2$\sigma$ level, as indicated previously), we derive the magnetic field angle by rotating the observed polarization angle (the electric field orientation, $\chi$), by 90$^{\circ}$. The top panel of Figure \ref{fig:bfield} shows the resulting magnetic field orientations displayed over the total intensity filaments. The distribution of magnetic field orientations are also presented as a histogram in the bottom panel of Figure \ref{fig:bfield}. In this histogram an angle of 0$^{\circ}$ indicates a magnetic field that is oriented parallel to the NTF orientation, with a clockwise increase in angle being positive and an anti-clockwise increase being negative.

\section{Discussion} \label{sec:disc}   

\subsection{Polarized Intensity Features} \label{sec:disc-pol}
As with previously observed NTFs, our ALMA observations reveal a discontinuous or ``patchy'' polarized intensity distribution that generally traces the NTF structure \citep[e.g.,][]{Gray1995,Lang1999a,Pare2019,Pare2024b}. We only observe significant polarized intensity in Southern declinations of the Radio Arc as can be seen in Figures \ref{fig:2panel} and \ref{fig:debiasedp}. The depolarization of the Radio Arc in its Northern extent has been observed in lower-frequency ($\sim$10 GHz) observations previously \citep[e.g.,][]{Yusef-Zadeh1987,Inoue1989}. This depolarization trend has been attributed to beam depolarization, line-of-sight contamination from molecular clouds coinciding with the Northern extent of the Radio Arc causing cancellation of the polarized signal of the Radio Arc, or the prominent \hii\ regions in the Northern region of the Radio Arc \citep{Yusef-Zadeh1988}. Since this depolarization is still observed at the 100 GHz ALMA observations presented here with higher resolution, we favor the possibilities that either the \hii\ regions or the complex molecular structures in the Northern portion of the Radio Arc could be producing the depolarization observed in this region of the NTF.

An additional distinct feature of the ALMA polarized intensity is that there is significant polarization that is not associated with significant total intensity. In particular, there is a coherent polarized intensity structure (labeled PF2 in Figure \ref{fig:2panel} and \ref{fig:debiasedp}) that is oriented parallel to the orientation of the Radio Arc filaments but lacks a significant total intensity counterpart.

To assess the origin of the polarization we note that previous VLA observations of the Radio Arc have revealed polarized intensity features that lack a significant total intensity counterpart \citep{Inoue1989,Pare2019}. These polarized intensity extensions are all observed to have fractional polarization p = P/I $\geq$1.0, indicative of missing total intensity as a result of the interferometric properties of the observations. We find an average fractional polarization of 1.4 for PF2. This result is non-physical, as one would not expect to obtain a fractional polarization $>$ 1.0; however,  this result supports the possibility that there is a total intensity counterpart to this polarization structure that is not sampled by the interferometric ALMA observations. The C43-3 ALMA array configuration has a maximum recoverable scale of $\sim$16.2\arcsec. Coherent structures much larger than this limit, such as the total intensity counterpart of PF2, are not being sampled. This conclusion is corroborated by the fact that PF1, which coincides with a total intensity NTF counterpart, has an average fractional polarization of 0.7.

Additionally, the large fractional polarization could result from polarized emission that originates from the central filamentary cores of the filaments. If this is the case, the emission from the outer layers of the filaments could have substantially lower polarization fraction since the magnetic field is less uniform through the thickness of the filament. This observational feature could therefore be a result of the polarized Stokes $Q$ and $U$ parameters having substantially different power spectra than the Stokes $I$, since $Q$ and $U$ depends on the polarization efficiency, the distribution of angles between the local magnetic fields and the line of sight, and the distribution of the magnetic field angles in the plane of the sky.

There is evidence supporting both of the above scenarios: previous VLA observations indicate that a radio shell of emission could be impacting the polarimetric distributions obtained for the Radio Arc \citep{Pare2019,Pare2021}. This brightness of the shell only appears on the limb of the shell projected in the plane of the sky because at an angular size of 10\arcmin it is a larger structure than the maximum recoverable scales of the VLA observations \citep{Simpson2007}. This same edge-brightening effect, however, does reveal the density enhancement within the shell. Observational evidence for the Radio Shell is also readily apparent in other interferometric observations of the region made using the VLA and MeerKAT \citep{Lang2010,Heywood2022}.

The radio shell has also been observed at infrared wavelengths \citep{Egan1998,Levine1999,Price2001,Simpson2007}, and these infrared observations reveal the structure to have a diameter of 10\arcmin. This shell encompasses the Quintuplet star cluster, which led \citet{Figer1999} to postulate that this shell could be a shock front generated by winds from this star cluster. \cite{Pare2021} argue that their Faraday effect analysis, coupled with these infrared observations, supports the possibility that the density enhancement of the shell alternately lies in the foreground and background along the line-of-sight with respect to the Radio Arc filaments.

However, the fact that the polarization of PF2 follows the NTF filamentary direction (as seen in Figure \ref{fig:debiasedp}) supports the possibility of the centrally concentrated polarization. Furthermore, the ALMA observations analyzed here have a higher resolution (pixel size of 0.3\arcsec) compared to the previous VLA observations ($\sim$1.0\arcsec), indicating that the ALMA observations could be resolving more of the fine-scale features of the Radio Arc polarization.

\subsection{Radio Arc Magnetic Field from ALMA}
\begin{figure*}
    \centering
    \includegraphics[width=1.0\textwidth]{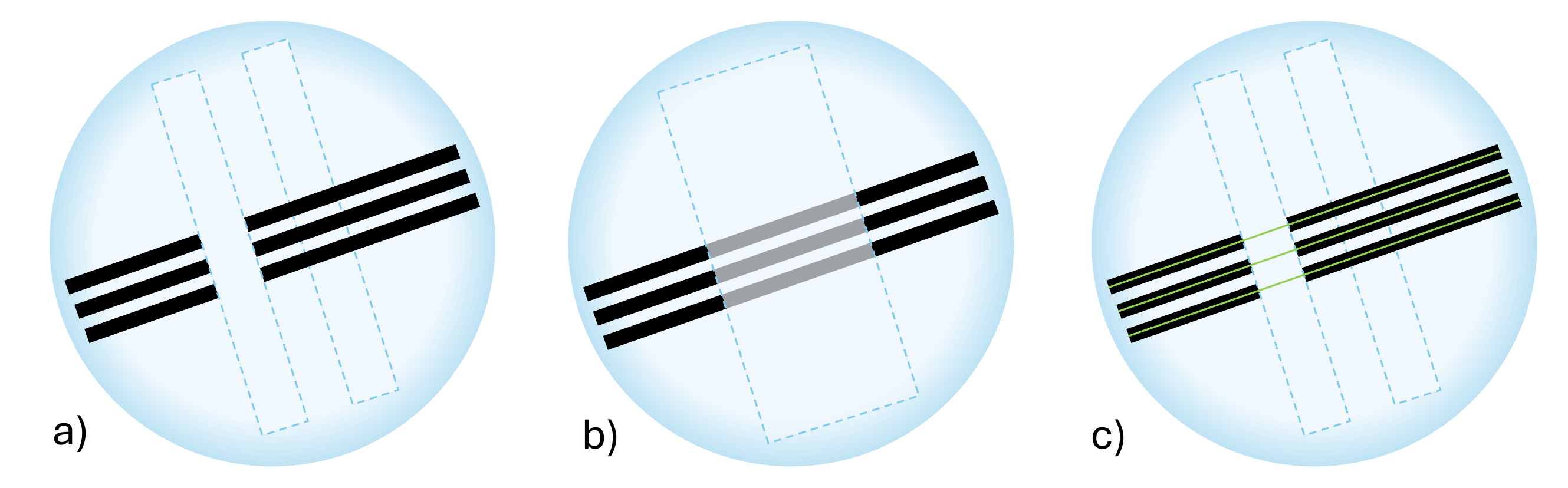}
    \caption{Sketches depicting possible line-of-sight geometries of structures local to the Radio Arc NTFs. The Radio Arc NTFs are shown as the solid black lines with the shall of radio emission represented as the enveloping circle. The left-hand panel (panel a) is a reproduction of the sketch from \citet{Pare2019} where the alternating magnetic field pattern inferred from their VLA data is thought to originate from density variations on the shell along different lines-of-sight towards the Radio Arc. In the middle panel (panel b) we present a sketch of the geometry with the assumption that the large fractional polarization inferred from our ALMA observations results from an un-sampled foreground structure. The right-hand panel (panel c) presents the sketch of the geometry with the assumption that the large fractional polarization results from the NTF filaments being strongly centrally polarized, where the green lines represent the central polarization.}
    \label{fig:sketch_new}
\end{figure*}
The observed magnetic field inferred from ALMA has two enhancements in orientation as seen in the lower panel of Figure \ref{fig:bfield}, with the enhancement for PF1 corresponding to angles ranging from 20 to 60$^{\circ}$ whereas the enhancement for PF2 corresponds to angles ranging from -80 to -20$^{\circ}$. We report an uncertainty on the magnetic field angles of $\pm0.29^{\circ}$ which was determined by propagating the uncertainties obtained on the Stokes $Q$ and $U$ distributions. The systematically different orientations between PF1 and PF2, coupled with the uniform field orientation within each filament, lies in stark contrast to what has been observed previously at lower frequencies ($\sim$10 GHz) where the magnetic field is predominantly parallel to the Radio Arc filament orientation \citep[e.g.,]{Yusef-Zadeh1987a,Inoue1989,Pare2019}. There are regions within the Radio Arc where a rotated magnetic field orientation was observed with $\sim$10 GHz VLA data \citep{Pare2019}. The locations of these rotated magnetic field regions as observed by the VLA are marked with blue rectangles in the top panel of Figure \ref{fig:bfield}. We find that the magnetic field orientations of PF1 and PF2 do not generally align with the rotated regions observed in the VLA.

The alternating magnetic field pattern observed from the VLA was argued to originate from electron density variations in a shell of radio emission that appears to envelop the Radio Arc \citep{Pare2019,Pare2021}, where the rotated magnetic field regions coincide with higher electron density regions of the radio shell that lie foreground to the Radio Arc along the line-of-sight. This geometry is depicted as a sketch in the left-hand panel of Figure \ref{fig:sketch_new} (panel a) where the Radio Arc filaments are shown as black lines and the radio shell is shown as the colored circle. The density enhancements are depicted as the dashed rectangles within the circle where one of the enhancements is foreground to the Radio Arc along the line-of-sight, resulting in a rotated magnetic field orientation. The other density enhancement is background to the NTFs, tracing the portion of the shell that is on the opposite side of the Radio Arc from our direction of observation. The background density enhancement therefore does not induce a rotated magnetic field geometry. This sketch was motivated by comparisons to the polarized intensity, rotation measure, and magnetic field distributions studied by \citet{Pare2019} and was subsequently corroborated using a more in-depth analysis of the line-of-sight Faraday effects encountered towards the source in \citet{Pare2021}.

In the middle panel of Figure \ref{fig:sketch_new} (panel b) we present one possible update to this sketch that is motivated by our 100 GHz ALMA polarimetric observations of the Radio Arc. In this updated sketch the radio shell has a more uniform density in all lines-of-sight, as represented by the single large dashed rectangle. In regions where the Radio Arc emission is bright along the line-of-sight, and is the dominant source of the polarized emission, the magnetic field is inferred from a combination of the polarization originating from the Radio Arc and the radio shell. Along lines-of-sight where the Radio Arc filaments are faint, however, the magnetic field is inferred from the polarization originating primarily from the radio shell. This explains the systematic difference in the magnetic field orientation observed between PF1 and PF2 in Figure \ref{fig:bfield}. A constant density in the shell also explains the uniform magnetic field orientation observed along the filament's lengths in PF1 and PF2.

We note, however, that the previous VLA observations also identified total intensity features that could possibly be density enhancements of the Radio Shell. These features are labeled in Figure 1 of \citet{Pare2019} as ``helical segments,'' and the coincidence of these features with the rotated magnetic field regions is used to support the sketch shown in the left-hand panel of Figure \ref{fig:sketch_new}. Indeed, if the large fractional polarizations of PF2 are a result of centrally concentrated polarization within the Radio Arc filaments, as discussed in Section \ref{sec:disc-pol}, then the uniform (but rotated) magnetic fields observed from these ALMA observations could indicate variations in the magnetic field between different Radio Arc filaments, rather than effects caused by a foreground intervening medium. This scenario is depicted in the right-hand panel of Figure \ref{fig:sketch_new} (panel c), where the central polarized cores are represented as green lines and the enveloping Radio Shell has the density enhancements inferred from the previous VLA observations \citep{Pare2019,Pare2021}.

We argue that panels b and c of Figure \ref{fig:sketch_new} are more accurate representations of the geometry local to the Radio Arc. Based on the ALMA observations presented here and the previous VLA observations, we slightly prefer the sketch shown in panel c of Figure \ref{fig:sketch_new} as the most accurate representation of the geometry local to the Radio Arc: the higher resolution ALMA observations are not as sensitive to the effects caused by the density enhancements of the Radio Shell, but are more sensitive to the potential of the polarization being centrally concentrated within the filaments. Conversely, the previous 10 GHz VLA observations are preferentially effected in the opposite manner, resulting in the alternating magnetic field pattern inferred from those observations \citep{Pare2019,Pare2021}.

Finally, we study whether there are any deformations in the Radio Arc that could help determine possible interaction between the NTFs and the Radio Shell. \citet{Tsuboi1997} use deformations in the Southernmost extent of the Radio Arc to infer interaction between the Radio Arc and a shell traced by CS 1 -- 0 and CS 2 -- 1 emission (this is a distinct shell from the one discussed in this present work). This Southern extent of the Radio Arc is outside the ALMA field of view, but we can inspect the observed portion of the Radio Arc for deformations to infer possible interaction. We do not see any such perturbations, but we note that this could be a result of the Radio Shell not having enough momentum to induce such perturbations given the high rigidity of the NTFs. Furthermore, the likely high magnetic field strengths of the Radio Arc NTFs (with estimates being $\sim$100s of $\mu$G) would mean that such perturbations would propagate quickly through the NTFs as Alfv\'en waves. The lack of deformations here can therefore not be invoked to conclusively rule out the possibility that these structures are interacting.

\section{Conclusions} \label{sec:conc}
In this work we have presented the first ever polarimetric ALMA observations of the unique GC NTFs which targeted the prominent Radio Arc. This also is the first time sub-arcsecond polarimetric observations have been conducted for the NTFs at 100 GHz. We derived the polarized intensity and magnetic field distributions for the Radio Arc from these observations, finding a magnetic field that is generally rotated with respect to the orientation of the Radio Arc NTFs but is uniform along the lengths of the Radio Arc filaments. These results build on previous VLA results showing an alternating magnetic field pattern for the Radio Arc where the field alternates between being parallel and rotated with respect to the Radio Arc orientation. We summarize the major findings of this work here:
\begin{itemize}
    \item The total intensity of the Radio Arc from ALMA (shown in the upper panel of Figure \ref{fig:2panel}) accurately depicts the Radio Arc filaments based on previous radio observations conducted with instruments like the VLA \citep[e.g.,][]{YMC1984,Pare2019,Heywood2022}.
    \item We find significant polarized intensity in the southern portion of the Radio Arc, with some of the polarization lacking a significant total intensity counterpart as shown in Figure \ref{fig:debiasedp} (labeled as PF2 in that figure). The features of this polarized intensity distribution are similar to previous radio polarimetric observations of the Radio Arc that reveal extensions of polarizations lacking total intensity filament counterparts \citep{Inoue1989,Pare2019}. These features are likely caused by a foreground structure that is coherent in total intensity on scales larger than the maximum recoverable scale of our interferometric observations.
    \item The magnetic field inferred from our ALMA observations lacks the alternating pattern observed from previous 10 GHz VLA observations \citep{Pare2019}. However, the field inferred from ALMA is ubiquitously rotated with respect to the Radio Arc filament orientation. The field orientation is also systematically different between PF1 and PF2 as shown in Figure \ref{fig:bfield}. 
    \item We update our hypothesis for the line-of-sight geometry of structures local to the Radio Arc based on the ALMA magnetic field results presented here, as shown in panels b and c of Figure \ref{fig:sketch_new}, building on the sketch presented in \citet{Pare2019}. We propose that sketch c is the most accurate depiction of reality, where the 100 GHz ALMA observations are most affected by the central concentration of polarization within the NTF filaments but are less sensitive to the Radio Shell density enhancements. This explains the different manifestations of the magnetic field distribution inferred for the VLA and ALMA observations.
\end{itemize}


\begin{acknowledgements}
    \textbf{Acknowledgments.}
    We would like to thank the anonymous reviewer for their helpful comments on this work. We would like to thank Dr. Natalie Butterfield (NRAO) for her insightful comments on the 3D orientation of the Radio Arc region. This paper makes use of the following ALMA data: ADS/JAO.ALMA\#2024.1.00515.S. ALMA is a partnership of ESO (representing its member states), NSF (USA) and NINS (Japan), together with NRC (Canada), NSTC and ASIAA (Taiwan), and KASI (Republic of Korea), in cooperation with the Republic of Chile. The Joint ALMA Observatory is operated by ESO, AUI/NRAO and NAOJ. The National Radio Astronomy Observatory is a facility of the National Science Foundation operated under cooperative agreement by Associated Universities, Inc. 
\end{acknowledgements}

\facility{
    ALMA
    }

\software{
    CASA,
    Astropy \citep{Astropy2022,Astropy2018,Astropy2013},
    Matplotlib \citep{Hunter2007}, 
    Numpy \citep{Harris2020},
    }

\bibliography{ALMA_RA}{}
\bibliographystyle{aasjournal}

\end{document}